\documentclass[aps,showpacs,amsmath,amssymb,prl,superscriptaddress,twocolumn]{revtex4}
\usepackage{graphicx}
\usepackage{epsfig}
\usepackage{ulem}
\usepackage{amsmath}
\usepackage{amssymb}
\usepackage{color}

\begin{document}

\title{Charge transfer along DNA dimers, trimers and polymers}
%\thanks{A footnote to the article title}

\author{Constantinos Simserides}
\email{csimseri@phys.uoa.gr}
\homepage{http://users.uoa.gr/~csimseri/}
\affiliation{National and Kapodistrian University of Athens, Faculty of Physics,
Panepistimiopolis, 15784 Zografos, Athens, Greece}
\date{\today}% It is always \today, today,

%\pacs{87.14.gk, 82.39.Jn, 82.39.Pj, 87.15.A-, 72.80.Le, 73.63.-b}

% 82.       Physical chemistry and chemical physics
% 82.39.-k 	Chemical kinetics in biological systems
% 82.39.Jn 	Charge (electron, proton) transfer in biological systems
% 82.39.Pj 	Nucleic acids, DNA and RNA bases

% 87. 	    Biological and medical physics
% 87.14.-g 	Biomolecules: types
% 87.14.gk 	DNA
% 87.15.-v 	Biomolecules: structure and physical properties
% 87.15.A- 	Theory, modeling, and computer simulation

% 72. 	    Electronic transport in condensed matter
% 72.80.-r 	Conductivity of specific materials
% 72.80.Le 	Polymers; organic compounds (including organic semiconductors)

% 73. 	    Electronic structure and electrical properties of surfaces, interfaces, thin films, and low-dimensional structures
% 73.63.-b 	Electronic transport in nanoscale materials and structures

\begin{abstract}
The transfer of electrons and holes along DNA dimers, trimers and polymers is described at the base-pair level, using the relevant on-site energies of the base-pairs and the hopping parameters between successive base-pairs.
The temporal and spatial evolution of carriers along a $N$ base-pair DNA segment is determined, solving a system of $N$
coupled differential equations.
Useful physical quantities are calculated including the \textit{pure} mean carrier transfer rate $k$,
the inverse decay length $\beta$ used for exponential fit ($k = k_0 \textrm{exp}(-\beta d)$) of the transfer rate as a function of the charge transfer distance $d = N \times$ 3.4 {\AA} and the exponent $\eta$ used for a power law fit ($k = k_0' N^{-\eta}$) of the transfer rate as function of the number of monomers $N$.
Among others, the electron and hole transfer along the polymers poly(dG)-poly(dC), poly(dA)-poly(dT), GCGCGC..., ATATAT... is studied.
%$\beta$ falls in the range $\approx$ 0.2 - 2 {\AA}$^{-1}$, $k_0$ is usually $\approx 10^{-2}$-10$^{-1}$ PHz although, generally, it falls in the wider range 10$^{-4}$-10 PHz.
%$\eta$ falls in the range $\approx$ 1.7 - 17, $k_0'$ is usually $\approx 10^{-2}$-10$^{-1}$ PHz, although generally, it falls in the wider range $\approx 10^{-4}$-10$^3$ PHz.
$\beta$ ($\eta$) falls in the range $\approx$ 0.2 - 2 {\AA}$^{-1}$ (1.7 - 17), $k_0$ ($k_0'$) is usually $\approx 10^{-2}$-10$^{-1}$ ($ 10^{-2}$-10$^{-1}$) PHz although, generally, it falls in the wider range $\approx 10^{-4}$-10 ($10^{-4}$-10$^3$) PHz.
The results are compared with past predictions and experiments.
%The results are compared with the predictions of Wang~\textit{et al.} Phys. Rev. Lett. \textbf{93}, (2004) 016401, as well as experiments, including
%Murphy~\textit{et al.} Science \textbf{262}, 1025 (1993);
%Arkin~\textit{et al.} Science \textbf{273}, 475 (1996);
%Giese~\textit{et al.} Angew. Chem. Int. Ed. \textbf{38}, 996 (1999);
%Giese~\textit{et al.} Nature \textbf{412}, 318 (2001).
Our approach illustrates to which extent a specific DNA segment can serve as an efficient medium for charge transfer.
\end{abstract}

\maketitle

%It is not the purpose of this letter to discuss why charge transfer along deoxyribonucleic acid (DNA) is crucial for molecular biology, genetics, and nanotechnology \cite{GenereuxBarton:2010, Giese:2002, Endres:2004}. Its aim is to present a convenient way to quantify electron or hole transfer along specific DNA segments using a simple tight-binding approach which can easily be implemented by interested colleagues.
Charge transfer along DNA is crucial for molecular biology, genetics, and nanotechnology \cite{GenereuxBarton:2010, Giese:2002, Endres:2004}. Here we present a convenient way to quantify electron or hole transfer along DNA segments using a tight-binding approach which can be easily implemented by interested colleagues.
To date all the tight-binding parameters relevant to charge transport along DNA either for electrons (traveling through LUMOs) or for holes (traveling through HOMOs) are available in the literature ~\cite{Endres:2004,HKS:2010-2011,Senthilkumar:2005,YanZhang:2002,SugiyamaSaito:1996,HutterClark:1996,ZhangLiEtAl:2002,
LiCaiSevilla:2001,LiCaiSevilla:2002,ShuklaLeszczynski:2002,Varsano:2006,Voityuk:2001,Migliore:2009,Kubar:2008,Ivanova:2008}.
Here we use them to study the temporal and spatial evolution of a carrier along DNA.
The transport of electrons or holes can be described at either
(I) the base-pair level or
(II) the single base level~\cite{HKS:2010-2011}.
We need the relevant on-site energies of either
(I) the base-pairs or
(II) the single bases.
In addition, we need the hopping parameters between either
(I) successive base-pairs or
(II) neighboring bases taking all possible combinations into account
[(IIa) successive bases in the same strand,
(IIb) complementary bases within a base-pair,
(IIc) diagonally located bases of successive base-pairs in opposite strands].
%Knowing those parameters,
To calculate the temporal and spatial evolution of carriers along a $N$ base-pair segment of DNA
one has to solve a system of either
(I) $N$ or
(II) $2N$
coupled differential equations.
Here we use the simplest approach (I) to examine charge transfer in B-DNA dimers, trimers and polymers.
%and to compare the results with available experiments.
Taking the relevant literature into account~\cite{Endres:2004,HKS:2010-2011,Senthilkumar:2005,YanZhang:2002,SugiyamaSaito:1996,HutterClark:1996,ZhangLiEtAl:2002,
LiCaiSevilla:2001,LiCaiSevilla:2002,ShuklaLeszczynski:2002,Varsano:2006,Voityuk:2001,Migliore:2009,Kubar:2008,Ivanova:2008},
we use the on-site energies and the hopping parameters shown in Tables~\ref{table:bpHL}-\ref{table:interbpTP}.
% in the supporting material.
We denote adenine (A), thymine (T), guanine (G), cytosine (C), and the relevant base-pairs A-T and G-C.
YX signifies two successive base-pairs:
the bases Y and X of two successive base-pairs (Y-Y$_{\textrm{compl}}$ and X-X$_{\textrm{compl}}$
separated and twisted by 3.4 {\AA} and $36^{\circ}$)
are located at the same strand in the direction $5'-3'$.
%For example, the notation AC denotes that the base-pair dimer consists of an
%adenine-thymine and a cytosine-guanine base-pair, where
%one strand contains A and C in the direction $5'-3'$ and the complementary
%strand contains T and G in the direction $3'-5'$.

For a description at the base-pair level,
the time-dependent single carrier (hole/electron)
wave function of the DNA segment of interest,
$\Psi^{DNA}_{H/L}({\bf r},t)$,
is considered as a linear combination of base-pair wave functions with
time-dependent coefficients,
%\begin{equation}\label{psi-total}
$\Psi^{DNA}_{H/L}({\bf r},t) = \sum_{\mu=1}^{N} A_{\mu}(t) \;
\Psi^{bp(\mu)}_{H/L}({\bf r})$.
%\end{equation}
$\Psi^{bp(\mu)}_{H/L}({\bf r})$ is the $\mu^{th}$
base-pair's HOMO or LUMO wave function ($H/L$).
The sum is extended over all base-pairs of the DNA segment under consideration.
$|A_{\mu}(t)|^2$ gives the probability of finding the carrier at base-pair $\mu$, at time $t$.
Starting from the time-dependent Schr\"{o}dinger equation,
%\begin{equation}\label{tdse}
$i \hbar \frac{\partial \Psi^{DNA}_{H/L}({\bf r},t)}{\partial t} =H^{DNA}\Psi^{DNA}_{H/L}({\bf r},t)$,
%\end{equation}
following the procedure % and the assumptions
described in Ref.~\cite{HKS:2010-2011},
we obtain that the time evolution of $A_\mu(t)$
obeys the tight-binding system of differential equations
\begin{equation}\label{TBbp}
i \hbar \frac{ dA_{\mu}} {dt} = E^{bp (\mu)}_{H/L} A_{\mu}
+ t^{bp (\mu;\mu-1)}_{H/L} A_{\mu-1}
+ t^{bp (\mu;\mu+1)}_{H/L} A_{\mu+1}.
\end{equation}
$E^{bp (\mu)}_{H/L}$ is the %HOMO/LUMO
on-site energy of base-pair $\mu$, and
$t^{bp (\mu; \mu ')}_{H/L}$ is the hopping parameter between base-pair $\mu$ and base-pair $\mu '$.
We can solve numerically the system of equations (\ref{TBbp})
and obtain,
through $A_{\mu}(t)$, the time evolution of a carrier propagating along the DNA segment of interest.

Regarding the tight-binding description of hole transport,
the corresponding tight-binding parameters should be taken with the opposite sign of
the calculated on-site energies and transfer hopping integrals~\cite{Senthilkumar:2005}.
This means that for describing hole transport at the base-pair level,
the on-site energies $E_H^{bp}$ presented in the second row of Table~\ref{table:bpHL} and
the hopping transfer integrals $t_H^{bp}$ presented in the second column of Table~\ref{table:interbpTP}
should be used with opposite signs to provide the tight-binding parameters of Eq.~\ref{TBbp}.
The on-site energies $E^{bp}_{H/L}$ for the two possible base-pairs A-T and G-C, calculated by various authors, are listed in Table~\ref{table:bpHL}. ${E^{bp \; \textrm{used}}_{H/L}}$ are the values actually used for the solution of Eq.~\ref{TBbp} in this article.
The hopping parameters $t_{H/L}^{bp}$ for all possible combinations of successive base-pairs, calculated by various authors, are given in Table~\ref{table:interbpTP}.
${t^{bp \; \textrm{used}}_{H/L}}$ are the values actually used for the solution of Eq.~\ref{TBbp} in this article.
Due to the symmetry between base-pair dimers YX and X$_{\textrm{compl}}$Y$_{\textrm{compl}}$,
the number of different hopping parameters is reduced from sixteen to ten.
In Table \ref{table:interbpTP} base-pair dimers
exhibiting the same transfer parameters are listed together in the first column.
We include in Table~\ref{table:interbpTP} the values listed:
in Table 3 of Ref.~\cite{HKS:2010-2011},
in Table II or Ref.~\cite{Voityuk:2001},
in Table 5 (``Best Estimates'') of Ref.~\cite{Migliore:2009},
in Table 4 of Ref.~\cite{Kubar:2008} (two estimations given),
in Table 2 of Ref.~\cite{Ivanova:2008}, and
the values extracted approximately from Fig.~4 of Ref.~\cite{Endres:2004}.
In Refs.~\cite{Kubar:2008,Migliore:2009,Ivanova:2008} all values given are positive,
in Ref.~\cite{Voityuk:2001} the authors explicitly state that they quote absolute values,
while in Refs.~\cite{HKS:2010-2011,Endres:2004} the sign is included.
In Ref.~\cite{HKS:2010-2011} all $t_H^{bp}$ and $t_L^{bp}$ have been calculated, while
in Ref.~\cite{Endres:2004} only the values of $t_H^{bp}$ for a few cases are approximately given.
According to Ref.~\cite{BlancafortVoityuk:2006} the approximation used in Ref.~\cite{Voityuk:2001} in general overestimates the transfer integrals. Summarizing, taking all the above into account, we use the values
${E^{bp \; \textrm{used}}_{H/L}}$ and ${t^{bp \; \textrm{used}}_{H/L}}$.

\begin{widetext}

\begin{table}[h!]
\caption{The on-site energies $E^{bp}_{H/L}$ for the two possible base-pairs A-T and G-C, calculated by various authors.
%{\color{red}${E^{bp \; \textrm{used}}_{H/L}}\mathrm{used}$}
${E^{bp \; \textrm{used}}_{H/L}}$
are the values actually used for the solution of Eq.~\ref{TBbp} in this article.
The first $\pi$-$\pi^*$ transition energies $E_{\pi-\pi^*}$ for the two B-DNA base-pairs are also shown.
Except for Ref.~\cite{HKS:2010-2011} these are \textit{ab initio} calculations which tend to overestimate the first $\pi$-$\pi^*$ transition energy.
All energies are given in eV.}
\centerline{
\begin{tabular}{|l|c|c|c|} \hline
B-DNA base-pair  & A-T     &   G-C   & reference             \\  \hline  \hline
$E_H^{bp}$       & $-$8.3  & $-$8.0  & \cite{HKS:2010-2011}  \\  \hline
$E_L^{bp}$       & $-$4.9  & $-$4.5  & \cite{HKS:2010-2011}  \\  \hline
$E_{\pi-\pi^*}$  &     3.4 &    3.5  & \cite{HKS:2010-2011}  \\  \hline
$E_H^{bp \; \mathrm{first \; pr.}}$     & $-$(7.8-8.2) & $-$(6.3-7.7) & \cite{SugiyamaSaito:1996,HutterClark:1996,ZhangLiEtAl:2002,LiCaiSevilla:2001,LiCaiSevilla:2002,ShuklaLeszczynski:2002} \\ \hline
$E_{\pi-\pi^*}^{\mathrm{first \; pr.}}$ &         6.4  &     4.3-6.3  & \cite{ShuklaLeszczynski:2002,Varsano:2006} \\ \hline
${E^{bp \; \textrm{used}}_{H}}$&   8.3 &    8.0 & \cite{HKS:2010-2011}  \\  \hline
${E^{bp \; \textrm{used}}_{L}}$&$-$4.9 & $-$4.5 & \cite{HKS:2010-2011}  \\  \hline
\end{tabular}  } \label{table:bpHL}
\end{table}

\begin{table}[h!]
\caption{The hopping parameters between successive base-pairs for all possible combinations. $t_H^{bp}$ ($t_L^{bp}$) refers to hole (electron) hopping through HOMOs (LUMOs). The notation is given in the text. The values listed
in Table 3 of Ref.~\cite{HKS:2010-2011},
in Table II or Ref.~\cite{Voityuk:2001},
in Table 5 (``Best Estimates'') of Ref.~\cite{Migliore:2009},
in Table 4 of Ref.~\cite{Kubar:2008} (two estimations given),
in Table 2 of Ref.~\cite{Ivanova:2008}, and
the values extracted approximately from Fig.~4 of Ref.~\cite{Endres:2004} are shown.
These quantities represent the parameters
$t^{bp (\mu;\mu\pm1)}_{H/L}$ which appear in Eq.~(\ref{TBbp}).
Finally, ${t^{bp \; \textrm{used}}_{H/L}}$ are the parameters actually used in this work
for the solution of Eq.~\ref{TBbp}. All hopping integrals $t_{H/L}^{bp}$ are given in meV.}
\centerline{
\begin{tabular}{|c|r|r|r|r|r|r|r||r|r|r|} \hline
Base-pair& $t_H^{bp}$ & $|t_H^{bp}|$ & $t_H^{bp}$ & $t_H^{bp}$ & $t_H^{bp}$ & $t_H^{bp}$ & $t_H^{bp \; \textrm{used}}$ & $t_L^{bp}$ & $t_L^{bp}$ & $t_L^{bp \; \textrm{used}}$ \\
sequence & \cite{HKS:2010-2011} & \cite{Voityuk:2001} & \cite{Endres:2004} & \cite{Migliore:2009} & \cite{Kubar:2008} & \cite{Ivanova:2008} & & \cite{HKS:2010-2011} & \cite{Endres:2004} &  \\ \hline \hline
AA, TT& $-$8& 26& $-$25 &8-17& 19(19)&22&    20  &$-$29& 35 & $-$29 \\ \hline
    AT&   20& 55&       &    & 47(74)&37& $-$35  &  0.5&    &     0.5  \\ \hline
AG, CT& $-$5& 25& $-$50 &    & 35(51)&43&    30  &    3& 35 &     3 \\ \hline
AC, GT&    2& 26&       &    & 25(38)&20& $-$10  &   32&    &    32 \\ \hline
    TA&   47& 50&       &    & 32(68)&52& $-$50  &    2&    &     2 \\ \hline
TG, CA& $-$4& 27&       &    & 11(11)&25&    10  &   17&    &    17 \\ \hline
TC, GA&$-$79&122&$-$160 &    &71(108)&60&   110  & $-$1& 35 &  $-$1 \\ \hline
GG, CC&$-$62& 93&$-$140 &  75&72(101)&63&   100  &   20& 35 &    20 \\ \hline
    GC&    1& 22&       &    & 20(32)&22& $-$10  &$-$10&    & $-$10 \\ \hline
    CG&$-$44& 78&       &    & 51(84)&74&    50  & $-$8&    &  $-$8 \\ \hline
\end{tabular} }   \label{table:interbpTP}
\end{table}

\end{widetext}

We define the column vector matrix $\vec{x}(t)$ made from  $A_j(t), \; j = 1, \dots, N$.
%\begin{equation}\label{x}
%$\vec{x}(t) = \left[
%\begin{array}{c}
%A_1(t) \\
%A_2(t) \\
%\vdots \\
%A_N(t)  \end{array} \right]$
%\end{equation}
Hence, %Eq.~~\ref{TBbp} reads
%\begin{equation}\label{xdotmathcalAx}
$\dot{\vec{x}}(t) = \widetilde{\mathcal{A}} \vec{x}(t)$,
%\end{equation}
%\begin{equation}\label{mathcalA}
$\widetilde{\mathcal{A}} = - \frac{i}{\hbar} \textrm{A}$.
%\end{equation}
%\begin{widetext}
%\begin{equation}\label{A}
%\textrm{A} = \left[
%\begin{array}{ccccccc}
%  E^{bp(1)}_{H/L}&t^{bp(1;2)}_{H/L}& 0                &\cdots&     0 &      0 & 0    \\
%t^{bp(2;1)}_{H/L}&  E^{bp(2)}_{H/L}& t^{bp(2;3)}_{H/L}&\cdots&     0 &      0 & 0    \\
%\vdots           & \vdots          & \vdots           &\vdots&\vdots & \vdots &\vdots\\
%0      & 0     & 0      &\cdots&t^{bp(N-1;N-2)}_{H/L}&  E^{bp(N-1)}_{H/L} & t^{bp(N-1;N)}_{H/L}    \\
%0      & 0     &  0     &\cdots&                   0 &t^{bp(N;N-1)}_{H/L} & E^{bp(N)}_{H/L} \end{array} \right].
%\end{equation}
%\end{widetext}
$\textrm{A}$ is a symmetric tridiagonal matrix. To proceed, %We solve Eq.~\ref{xdotmathcalAx} using the
we use the
{\it eigenvalue method}, i.e. we look for solutions of the form $\vec{x}(t) = \vec{v} e^{\tilde{\lambda} t} \Rightarrow \dot{\vec{x}}(t) = \tilde{\lambda} \vec{v} e^{\tilde{\lambda} t}$. %Hence,
%Eq.~\ref{xdotmathcalAx} reads
%\begin{equation}\label{mathcalA}
$\widetilde{\mathcal{A}} \vec{v} = \tilde{\lambda} \vec{v}$,
%\end{equation}
or
%\begin{equation}\label{textrmA}
$\textrm{A} \vec{v} = \lambda \vec{v}$,
%\end{equation}
with $ \tilde{\lambda} = - \frac{i}{\hbar} \lambda $.
Having checked that the normalized eigenvectors $\vec{v_k}$ corresponding to the eigenvalues $\lambda_k$ % of Eq.~\ref{textrmA}
are linearly independent, the solution
%of our problem
is
%\begin{equation}
$\vec{x}(t) = \sum_{k=1}^{N} c_k \vec{v_k} e^{-\frac{i}{\hbar} \lambda_k t}$.
%\end{equation}
%The initial conditions used usually in this article are
%\begin{equation}\label{x0}
%\vec{x}(0) = \left[
%\begin{array}{c}
%A_1(0) \\
%A_2(0) \\
%\vdots \\
%A_N(0)  \end{array} \right]
%=
%\left[
%\begin{array}{c}
%1 \\
%0 \\
%\vdots \\
%0  \end{array} \right]
%\end{equation}
%which means that
%We initially place the carrier in base-pair 1 and we want to see how the carrier will evolve, time passing.
From the initial conditions we determine $c_i(t)$.
%In some cases, however, comparing with a particular experiment, we need to put the carrier initially at the base-pair indicated by the authors of the experimental work.

%We emphasize that our calculations below refer to \textit{pure} carrier transfer rates extracted from the probabilities to find the carrier at a %particular monomer of interest after having placed it initially (for time zero) at another monomer, i.e. refer directly to the solution of %Eq.~(\ref{TBbp}).
%%These are extracted from formulas of the type given in Eq.~(\ref{meantransferrateN}) or the relevant for dimers and trimers %Eq.~(\ref{meantransferrate2}) and Eq.~(\ref{meantransferrate3}) below.
%We do not take into account the influence of other factors such as the density of states, the environment etc.

\begin{figure*}[]
\hspace{0cm}
\centering
\includegraphics[width=7cm]{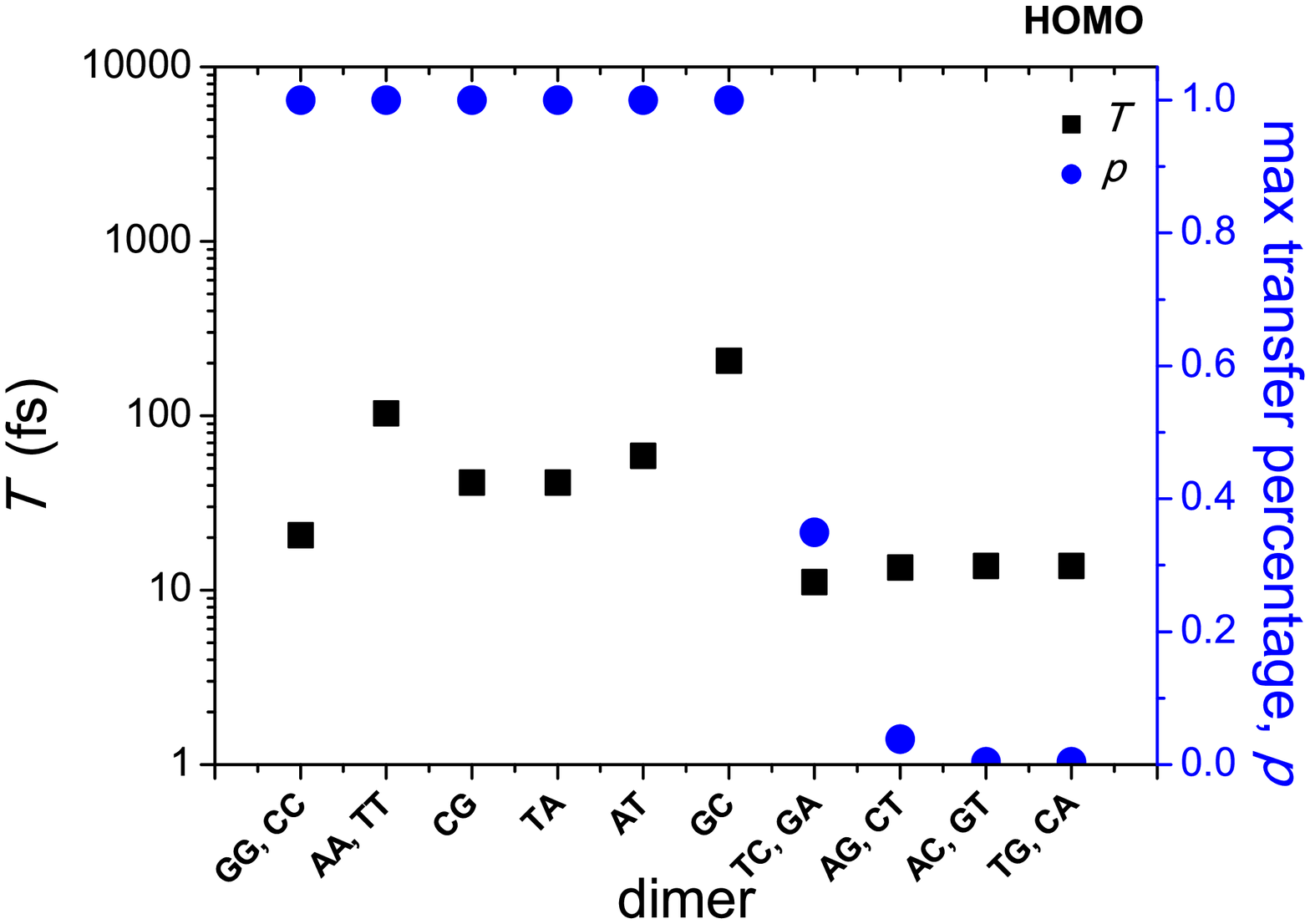}
\includegraphics[width=7cm]{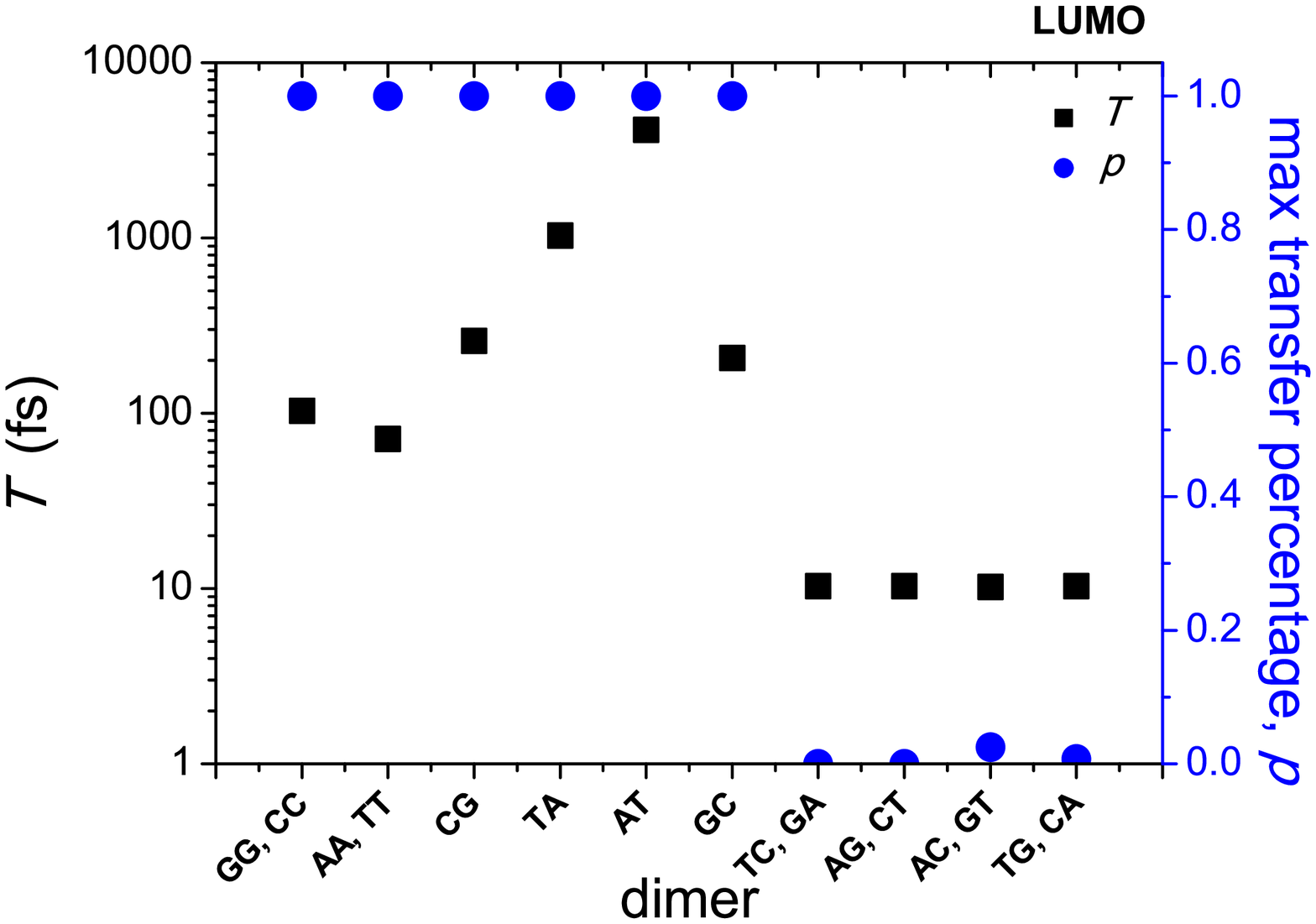}
\includegraphics[width=7cm]{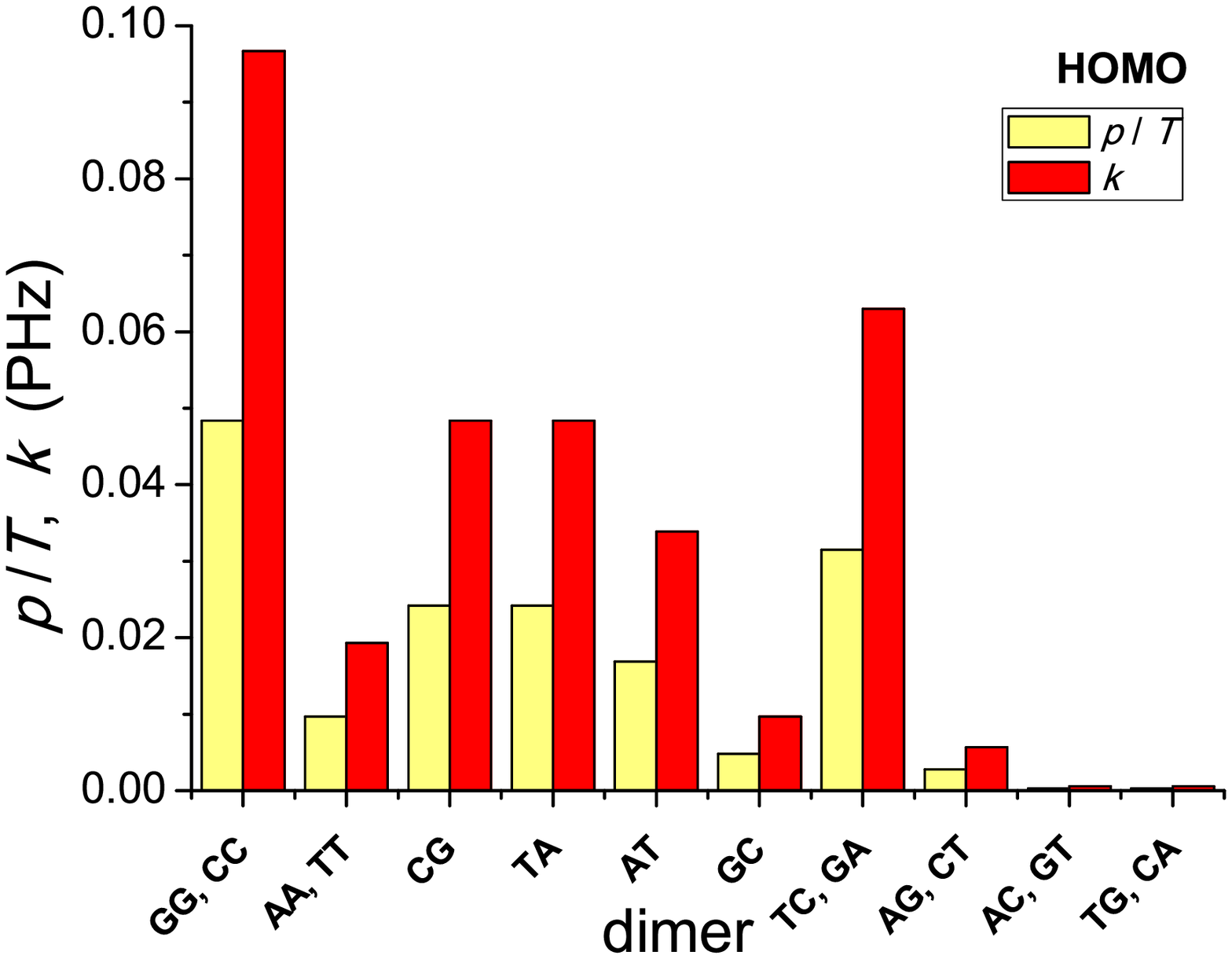}
\includegraphics[width=7cm]{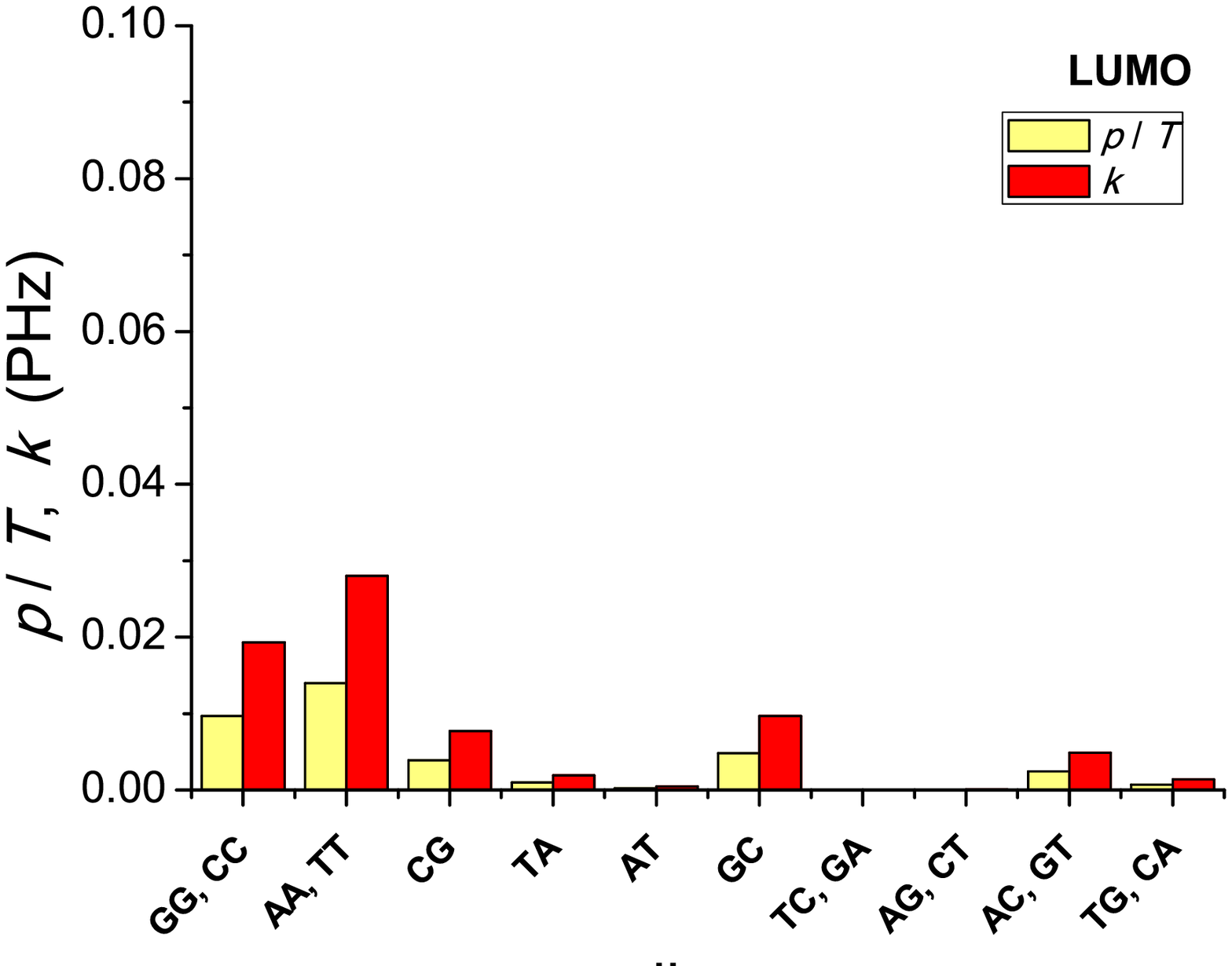}
\includegraphics[width=7cm]{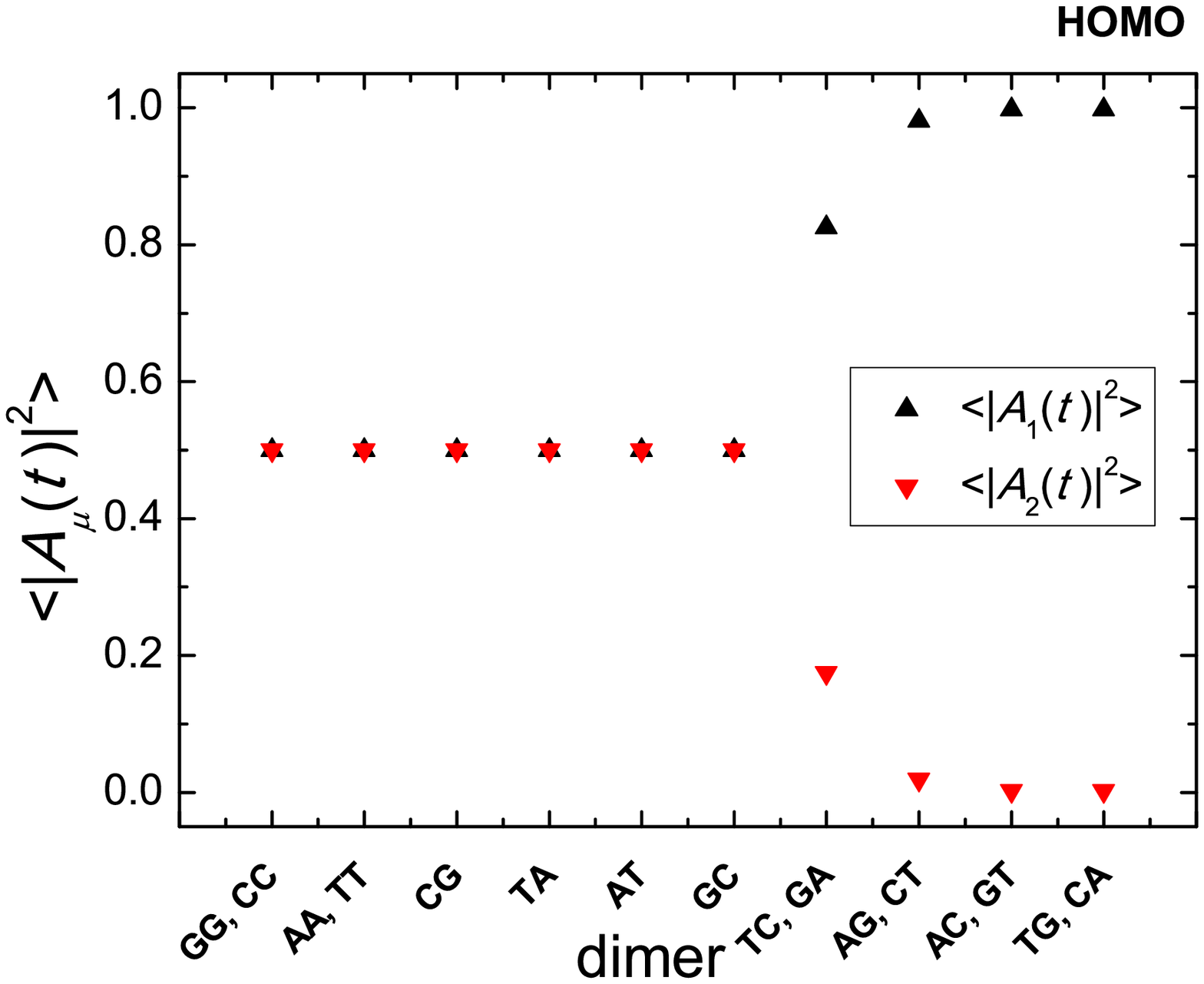}
\includegraphics[width=7cm]{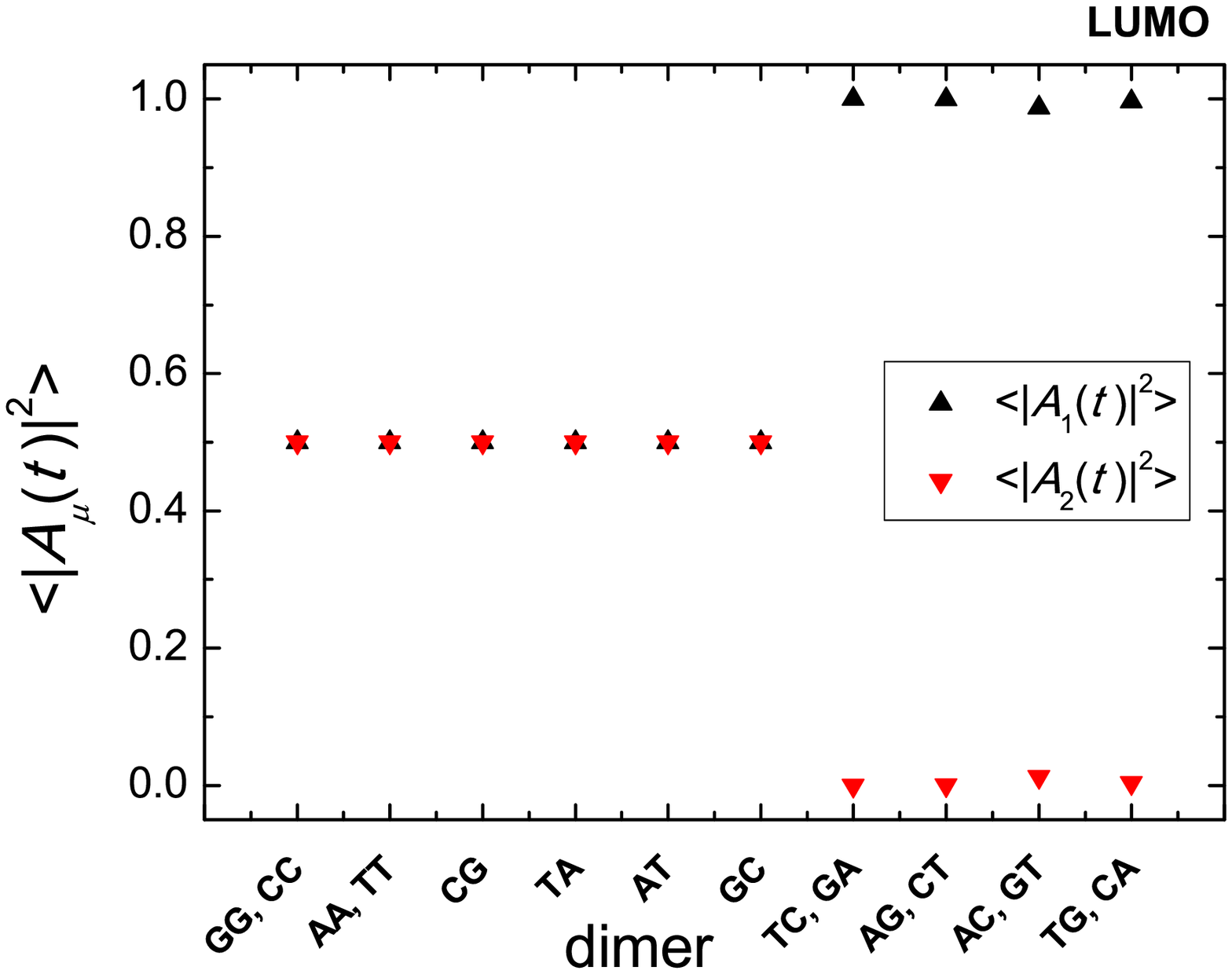}
\caption{Dimers:
the period of carrier transfer between monomers $T$ (fs) and the maximum transfer percentage $p$ [1st row],
the \textit{pure} maximum transfer rate defined as $p/T$ (PHz) and
the \textit{pure} mean transfer rate defined as $k = \langle |A_{2}(t)|^2 \rangle / {t_{2}}_{mean} $ (PHz) [2nd row].
$\langle|A_{\mu}(t)|^2\rangle, \; \mu = 1, 2$, which describe the spread of the carrier over the monomers constituting the dimer [3rd row].
For the dimers made up of identical monomers $p=1$ whereas for the dimers made up of different monomers $p<1$.
In the latter case, the \textit{pure} maximum transfer rate and the \textit{pure} mean transfer rate are negligible
for HOMO hole transfer when purines are crosswise to pyrimidines (GT $\equiv$ AC and CA $\equiv$ TG dimers) and
for LUMO electron transfer when purines are on top of pyrimidines (GA $\equiv$ TC and CT $\equiv$ AG dimers).
For dimers $k = 2 \frac{p}{T}$.}
\label{dimersholeselectrons}
\end{figure*}

For \emph{dimers}, supposing that $\lambda_2 \ge \lambda_1$,
%the solution of our problem is
%\begin{equation}\label{dimersolution}
%\vec{x}(t) = \sum_{k=1}^{2} c_k \vec{v_k} e^{-\frac{i}{\hbar} \lambda_k t}.
%\end{equation}
%Let us suppose that $\lambda_2 \ge \lambda_1$.
%We are interested in the quantities $|A_{\mu}(t)|^2, \mu = 1, 2$ since they provide the probabilities of finding the carrier at the base-pair $\mu$. %From Eq.~\ref{dimersolution},
we obtain the period of $|A_{\mu}(t)|^2, \mu = 1, 2$,
%\begin{equation}\label{period}
$T = \frac{h}{\lambda_2 - \lambda_1}$.
%\end{equation}
For a dimer consisting of two identical monomers with purine on purine %and pyrimidine on pyrimidine,
(GG $\equiv$ CC, AA $\equiv$ TT),
%Maybe in the simplest case one could imagine, suppose that we have a dimer consisting of two identical monomers with purine on purine and pyrimidine on pyrimidine, i.e. GG $\equiv$ CC or AA $\equiv$ TT
%e.g.
%\begin{eqnarray}
%5' &      & 3' \nonumber \\
%\textrm{G}  &   -  & \textrm{C} \nonumber \\
%\textrm{G}  &   -  & \textrm{C} \nonumber  \\
%3' &      & 5'
%\label{GG}
%\end{eqnarray}
%so that
%\begin{equation}\label{Auniform}
%\textrm{A} = \left[
%\begin{array}{cc}
%E^{bp} & t^{bp}  \\
%t^{bp} & E^{bp}  \end{array} \right]
%\end{equation}
%with eigenvalues
%\begin{equation}\label{EigenvIdMonomers}
$\lambda_{1,2} = E^{bp} \mp t^{bp}$.
%\end{equation}
%and corresponding normalized eigenvectors
%\begin{equation}
%\vec{v_1} = \left[
%\begin{array}{c}
%-\sqrt{2}/2 \\
%\sqrt{2}/2 \end{array} \right]
%,
%\vec{v_2} = \left[
%\begin{array}{c}
%\sqrt{2}/2 \\
%\sqrt{2}/2 \end{array} \right].
%\end{equation}
Then, if we initially place the carrier in monomer 1,
%Then, for initial condition
%\begin{equation}
%\vec{x}(0) =
%\left[
%\begin{array}{c}
%1 \\
%0 \end{array} \right],
%\end{equation}
%\begin{equation}\label{vecxt2}
%\vec{x}(t) \!\!=\!\!
%\left[
%\begin{array}{c}
%A_1(t) \\
%A_2(t) \end{array} \right]
%\!\!=\!\!
%\left[
%\begin{array}{r}
% \frac{1}{2} e^{-\frac{i}{\hbar} \lambda_1 t} + \frac{1}{2} e^{-\frac{i}{\hbar} \lambda_2 t}    \\
%-\frac{1}{2} e^{-\frac{i}{\hbar} \lambda_1 t} + \frac{1}{2} e^{-\frac{i}{\hbar} \lambda_2 t} \end{array} \right].
%\end{equation}
%To the eigenvalues $\lambda_i$ ($i = 1, 2$) of Eq.~\ref{vecxt2} correspond the frequencies $f_i$ and the periods $T_i$ defined by  $\frac{\lambda_i}{\hbar} = \frac{2\pi}{T_i} = 2 \pi f_i$, i.e.
%\begin{equation}
%\frac{f_2}{f_1} = \frac{T_1}{T_2} = \frac{\lambda_2}{\lambda_1} = \frac{E^{bp} + t^{bp}}{E^{bp} - t^{bp}}.
%\end{equation}
%In reality, as mentioned above,
%We are interested in the quantities $|A_{\mu}(t)|^2, \mu = 1, 2$ since they provide the probabilities of finding the carrier at the base-pair $\mu$. From Eq.~\ref{vecxt2} we obtain
%\begin{equation}\label{vecxt2squared}
%\left[
%\begin{array}{c}
%|A_1(t)|^2 \\
%|A_2(t)|^2 \end{array} \right]
%=
%\left[
%\begin{array}{r}
%\frac{1}{2} + \frac{1}{2}\textrm{cos}[\frac{(\lambda_2 - \lambda_1)t}{\hbar}] \\
%\frac{1}{2} - \frac{1}{2}\textrm{cos}[\frac{(\lambda_2 - \lambda_1)t}{\hbar}] \end{array} \right].
%\end{equation}
$|A_1(t)|^2 = \frac{1}{2} + \frac{1}{2}\textrm{cos}[\frac{(\lambda_2 - \lambda_1)t}{\hbar}]$,
$|A_2(t)|^2 = \frac{1}{2} - \frac{1}{2}\textrm{cos}[\frac{(\lambda_2 - \lambda_1)t}{\hbar}]$.
%Next, suppose that we have
For a dimer consisting of two identical monomers with purine on pyrimidine %and vice versa,
(GC, CG, AT, TA),
%e.g.
%\begin{eqnarray}
%5' &      & 3' \nonumber \\
%\textrm{G}  &   -  & \textrm{C} \nonumber \\
%\textrm{C}  &   -  & \textrm{G} \nonumber  \\
%3' &      & 5'
%\label{GC}
%\end{eqnarray}
%In these cases the matrix $\textrm{A}$ is given again by Eq.~\ref{Auniform} i.e.
the problem is identical. %to the previous one.
%Suppose now that we have
For a dimer made up of different monomers (AG \!\!\! $\equiv$ \!\!\! CT, AC \!\!\! $\equiv$ \!\!\! GT, TG \!\!\! $\equiv$ \!\!\! CA,  TC \!\!\! $\equiv$ \!\!\! GA),
%e.g.
%\begin{eqnarray}
%5' &      & 3' \nonumber \\
%\textrm{G}  &   -  & \textrm{C} \nonumber \\
%\textrm{A}  &   -  & \textrm{T} \nonumber  \\
%3' &      & 5'.
%\label{GG}
%\end{eqnarray}
%Then,
%\begin{equation}\label{GAorTC}
%\textrm{A} = \left[
%\begin{array}{cc}
%E^{bp1} & t^{bp}  \\
%t^{bp} & E^{bp2}  \end{array} \right]
%\end{equation}
%with eigenvalues
%\begin{equation}\label{EigenvDiMonomers}
$\lambda_{1,2} = \frac{E^{bp1} + E^{bp2}}{2} \mp \sqrt{\frac{(E^{bp1} - E^{bp2})^2}{4}+{t^{bp}}^2}$.
%\end{equation}
%Eq.~\ref{EigenvIdMonomers} is a special case of Eq.~\ref{EigenvDiMonomers} for $E^{bp1} = E^{bp2}$.
%If the dimer is made up of identical monomers (different monomers), the eigenvalues are given by Eq.~\ref{EigenvIdMonomers} (Eq.~\ref{EigenvDiMonomers}).
%We depict this schematically in Fig.~\ref{dimers}.
%Let us define $\Delta^{bp} = |E^{bp1} - E^{bp2}|$.
%In the special case $\lambda_2 = \lambda_1 \Leftrightarrow [t^{bp} = 0 \textrm{ and } \Delta^{bp} = 0]$ (cf. Eq.~\ref{EigenvIdMonomers} and Eq.~\ref{EigenvDiMonomers}), $T\to \infty$, i.e. the functions are constant.
%Hence,
%Using Eq.~\ref{EigenvIdMonomers}
Hence, for identical monomers
%\begin{equation}\label{periodIdMonomers}
$T \!\! = \!\! \frac{h}{2 |t^{bp}|}$,
%\end{equation}
%and
%Using Eq.~\ref{EigenvDiMonomers}
for different monomers,
%\begin{equation}\label{perioddimers}
$T = \frac{h}{\sqrt{(2t^{bp})^2 + (\Delta^{bp})^2}}$.
%\end{equation}
%The former Eq.~\ref{periodIdMonomers} is a special case of the latter Eq.~\ref{perioddimers} for $E^{bp1} = E^{bp2}$.
%Hence, in the case of identical monomers the period depends only on the hopping parameter between the base-pairs, but in the case of different monomers the period depends {\it additionally} on the energy gap between the on-site energies of the carrier in the different monomers.
$\Delta^{bp} = |E^{bp1} - E^{bp2}|$.
%From Eq.~\ref{dimersolution} we obtain the
The maximum transfer percentage of the carrier from base-pair 1 to base-pair 2,
$p = 4 c_1 v_{11} c_2 v_{12}$. This refers to the maximum of $|A_{2}(t)|^2$.
$v_{ij}$ is the $i$-th component of eigenvector $j$.
%The maximum transfer percentage reads
Hence,
%\begin{equation}\label{pdimers}
$p = \frac{(2t^{bp})^2}{(2t^{bp})^2+(\Delta^{bp})^2}$.
%\end{equation}
For identical (different) monomers, $p = 1$ ($p < 1$).
The \textit{pure} maximum transfer rate
%({\bf maybe not the best name}),
can be defined as
%\begin{equation}\label{pperTdimers}
$\frac{p}{T} = \frac{(2t^{bp})^2}{h\sqrt{(2t^{bp})^2+(\Delta^{bp})^2}}$.
%\end{equation}
For identical monomers, %it follows that
%\begin{equation}\label{pperTIdMonomers}
$\frac{p}{T} = \frac{2|t^{bp}|}{h}$.
%\end{equation}
%For a dimer consisting of two identical monomers with purine on purine and pyrimidine on pyrimidine, as mentioned above, the period of $|A_{\mu}(t)|^2, \mu = 1, 2$ is given by Eq.~\ref{periodIdMonomers}. In all cases, the maximum transfer percentage, $ p = 1$ (100\%).
%For a dimer consisting of two identical monomers with purine on pyrimidine and vice versa, again, we use Eq.~\ref{periodIdMonomers}. In all cases, the maximum transfer percentage, $ p = 1$ (100\%). Conclusively,
%For all cases of a dimer consisting of two identical monomers, $ p = 1$ (100\%).
%In contrast, for a dimer made up of two different monomers $ p < 1$.
For holes, when purines are crosswise to pyrimidines (GT $\equiv$ AC, CA $\equiv$ TG) $p$ is negligible, hence, we expect that insertion of these dimers in a sequence of DNA base-pairs will disrupt {\it hole} transfer.
Also AG $\equiv$ CT has very small $p$.
%For electrons, again, we observe that in contrast to the case of identical monomers $ p < 1$.
%Additionally,
Generally, electrons have smaller $p$ than holes.
In contrast to the cases of holes, %mentioned just above,
when purines are NOT crosswise to pyrimidines (GA $\equiv$ TC, CT $\equiv$ AG) $p$ is negligible, hence, we expect that insertion of these dimers in a sequence of DNA base-pairs will disrupt {\it electron} transfer.
% an hypothesis that has to be checked later on.  [$\maltese$].
Generally, in cases of different monomers $T$ is smaller than
in cases of identical monomers due to
%the difference between Eq.~\ref{perioddimers} and Eq.~\ref{periodIdMonomers}, i.e.
the extra term containing $\Delta^{bp} = |E^{bp1} - E^{bp2}|$.
Overall, carrier transfer is more difficult for different monomers compared to identical monomers.
If $|A_{2}(0)|^2 = 0$, a \textit{pure} mean transfer rate can be defined as
%\begin{equation}\label{meantransferrate2}
$k = \frac{\langle |A_{2}(t)|^2 \rangle}{{t_{2}}_{mean}}$,
%\end{equation}
where ${t_{2}}_{mean}$ is the first time $|A_{2}(t)|^2$ becomes equal to $\langle |A_{2}(t)|^2 \rangle$ i.e.
``the mean transfer time''.
Figure~\ref{dimersholeselectrons} shows % quantities relevant to the periodic carrier transfer in a base-pair dimer, specifically,
%the hopping integrals $t_{H/L}^{bp}$,
%the period of carrier transfer between monomers
$T$,
%and the maximum transfer percentage
$p$,
%the \textit{pure} maximum transfer rate defined as
$p/T$ and
%the \textit{pure} mean transfer rate defined as
$k = \langle |A_{2}(t)|^2 \rangle / {t_{2}}_{mean} $.
% as well as $\langle|A_{\mu}(t)|^2\rangle, \; \mu = 1, 2$, for all dimers, both for electrons and holes.
%which describe the spread of the carrier over the monomers constituting the dimer.
%For the dimers made up of identical monomers $p=1$ whereas for the dimers made up of different monomers $p<1$.
%In the latter case, the \textit{pure} maximum transfer rate and the \textit{pure} mean transfer rate are negligible for HOMO hole transfer when purines are crosswise to pyrimidines (GT $\equiv$ AC and CA $\equiv$ TG dimers) and for LUMO electron transfer when purines are on top of pyrimidines (GA $\equiv$ TC and CT $\equiv$ AG dimers). For dimers $k = 2 \frac{p}{T}$.

For \emph{trimers}, supposing that $\lambda_1 \le \lambda_2 \le \lambda_3$,
%For any trimer, the solution of our problem is
%\begin{equation}\label{trimersolution}
%\vec{x}(t) = \sum_{k=1}^{3} c_k \vec{v_k} e^{-\frac{i}{\hbar} \lambda_k t}.
%\end{equation}
%Let us suppose further that $\lambda_1 \le \lambda_2 \le \lambda_3$.
%We are interested in the quantities $|A_{\mu}(t)|^2, \mu = 1, 2, 3$ %(cf. Eq.~\ref{psi-total} and Eq.~\ref{x})  since they provide the probabilities of finding the carrier at the base-pair $\mu$. From Eq.~\ref{trimersolution},
we conclude that $|A_{\mu}(t)|^2, \mu = 1, 2, 3$ are sums of terms containing constants and periodic functions with periods
%\begin{equation}\label{periods}
$T_{21} = \frac{h}{\lambda_2 - \lambda_1},
T_{32} = \frac{h}{\lambda_3 - \lambda_2},
T_{31} = \frac{h}{\lambda_3 - \lambda_1}$.
%\end{equation}
%In case of double degeneracy, e.g. $\lambda_2 = \lambda_1 \Rightarrow T_{21} \to \infty$, i.e. the terms containing $T_{21}$ are constant and the only period that remains is $T_{32} = T_{31}$, hence $|A_{\mu}(t)|^2, \mu = 1, 2, 3$ are periodic. In case of triple degeneracy, $\lambda_3 = \lambda_2 = \lambda_1$ all periods are  infinite, hence $|A_{\mu}(t)|^2, \mu = 1, 2, 3$ are constants.
There are 8 trimers consisting of identical monomers.
%4 based on G-C monomers [GGG $\equiv$ CCC (0 times crosswise purines), GGC $\equiv$ GCC (1 time crosswise purines), CGG $\equiv$ CCG (1 time crosswise %purines), GCG $\equiv$ CGC (2 times crosswise purines)], and
%4 based on A-T monomers [AAA $\equiv$ TTT (0 times crosswise purines), AAT $\equiv$ ATT (1 time crosswise purines), TAA $\equiv$ TTA (1 time crosswise %purines), ATA $\equiv$ TAT (2 times crosswise purines)].
In the cases of 0 times crosswise purines
%\begin{equation}\label{ATrimersIdMonomers0}
%\textrm{A} = \left[
%\begin{array}{ccc}
%E^{bp} & t^{bp} & 0      \\
%t^{bp} & E^{bp} & t^{bp} \\
%0      & t^{bp} & E^{bp} \end{array} \right]
%\end{equation}
%with eigenvalues
%\begin{equation}\label{ETrimersIdMonomers0}
$\lambda_2 = E^{bp},
\lambda_{1,3} = E^{bp} \mp t^{bp} \sqrt{2}$.
%\end{equation}
Hence, two periods are involved in $|A_{\mu}(t)|^2, \mu = 1, 2, 3$:
%\begin{equation}\label{periodsTrimersIdMonomers0}
$T_{M} = \frac{h}{t^{bp} \sqrt{2}},
T_{E} = \frac{h}{2 t^{bp} \sqrt{2}}
\Rightarrow \;\; \frac{T_{M}}{T_{E}} = \frac{2}{1}$.
%\end{equation}
$T_{M}=T_{M(21)}=T_{M(32)}$ involves the Medium eigenvalue, $T_{E}=T_{E(31)}$ involves only the Edge eigenvalues.
Since $\frac{T_{M}}{T_{E}} = \frac{2}{1}$, $|A_{\mu}(t)|^2, \mu = 1, 2, 3$ are periodic.
In the cases of 1 or 2 times crosswise purines
%\begin{equation}\label{ATrimersIdMonomers12}
%\textrm{A} = \left[
%\begin{array}{ccc}
%E^{bp} & t^{bp}  & 0      \\
%t^{bp} & E^{bp}  & t^{bp'} \\
%0      & t^{bp'} & E^{bp} \end{array} \right]
%\end{equation}
%with eigenvalues
%\begin{equation}\label{ETrimersIdMonomers12}
$\lambda_2 = E^{bp},
\lambda_{1,3} = E^{bp} \mp \sqrt{{t^{bp}}^2+{t^{bp'}}^2}$.
%\end{equation}
Hence, two periods are involved in $|A_{\mu}(t)|^2, \mu = 1, 2, 3$:
%\begin{equation}\label{periodsTrimersIdMonomers12}
$T_{M} = \frac{h}{\sqrt{{t^{bp}}^2+{t^{bp'}}^2}},
T_{E} = \frac{h}{2 \sqrt{{t^{bp}}^2+{t^{bp'}}^2}}
\Rightarrow \frac{T_{M}}{T_{E}} = \frac{2}{1}$.
%\end{equation}
Since $\frac{T_{M}}{T_{E}} = \frac{2}{1}$ it follows that $|A_{\mu}(t)|^2, \mu = 1, 2, 3$ are periodic.
Conclusively, in all cases of a trimer consisting of identical monomers, $|A_{\mu}(t)|^2, \mu = 1, 2, 3$ are periodic with period $T_{M}$.
Suppose that we have a trimer consisting of different monomers. There are 24 different such trimers.
%GGA $\equiv$ TCC, GGT $\equiv$ ACC, GCA $\equiv$ TGC, GCT $\equiv$ AGC, GAG $\equiv$ CTC, GAC $\equiv$ GTC,
%GAA $\equiv$ TTC, GAT $\equiv$ ATC, GTG $\equiv$ CAC, GTA $\equiv$ TAC, GTT $\equiv$ AAC, CGA $\equiv$ TCG,
%CGT $\equiv$ ACG, CCA $\equiv$ TGG, CCT $\equiv$ AGG, CAG $\equiv$ CTG, CAA $\equiv$ TTG, CAT $\equiv$ ATG,
%CTA $\equiv$ TAG, AGA $\equiv$ TCT, AGT $\equiv$ ACT, ACA $\equiv$ TGT, TGA $\equiv$ TCA, CTT $\equiv$ AAG.
For example, suppose that we refer to HOMO charge transfer in GAC $\equiv$ GTC, then
%\begin{equation}\label{ATrimersDiMonomersGAC}
%\textrm{A} = \left[
%\begin{array}{ccc}
%E^{bp} & t^{bp}  & 0      \\
%t^{bp} & E^{bp''}  & t^{bp'} \\
%0      & t^{bp'} & E^{bp} \end{array} \right]
%\end{equation}
%where $E^{bp''} > E^{bp}$, and the eigenvalues are
with $E^{bp''} > E^{bp}$,
%\begin{eqnarray}
$\lambda_2 = E^{bp},
\lambda_{1,3} \! = \! \frac{E^{bp}+E^{bp''}}{2} \! \mp \!
\sqrt{\left(\frac{E^{bp}-E^{bp''}}{2}\right)^2 \! + \! {t^{bp}}^2+{t^{bp'}}^2 }$.
%\label{ETrimersIdMonomers12}
%\end{eqnarray}
Three periods are involved in $|A_{\mu}(t)|^2, \mu = 1, 2, 3$.
With $\Delta^{bp} = |E^{bp} - E^{bp''}|$,
%\begin{eqnarray}\label{periodsTrimersIdMonomers12}
$T_{M(32)} = \frac{h}{\frac{\Delta^{bp}}{2} +
\sqrt{\frac{{\Delta^{bp}}^2}{4} + {t^{bp}}^2+{t^{bp'}}^2 }},
T_{E(31)} = \frac{h}{2 \sqrt{\frac{{\Delta^{bp}}^2}{4} + {t^{bp}}^2+{t^{bp'}}^2 }},
T_{M(21)} = \frac{h}{-\frac{\Delta^{bp}}{2} +
\sqrt{\frac{{\Delta^{bp}}^2}{4} + {t^{bp}}^2+{t^{bp'}}^2 }}.$
%\end{eqnarray}
%Therefore,
$\frac{T_{M(32)}}{T_{E(31)}}$ and $\frac{T_{M(21)}}{T_{E(31)}}$ may be irrational numbers, hence  $|A_{\mu}(t)|^2, \mu = 1, 2, 3$ may be non-periodic.
\begin{figure}[]
\hspace{0cm}
\centering
\includegraphics[width=8cm]{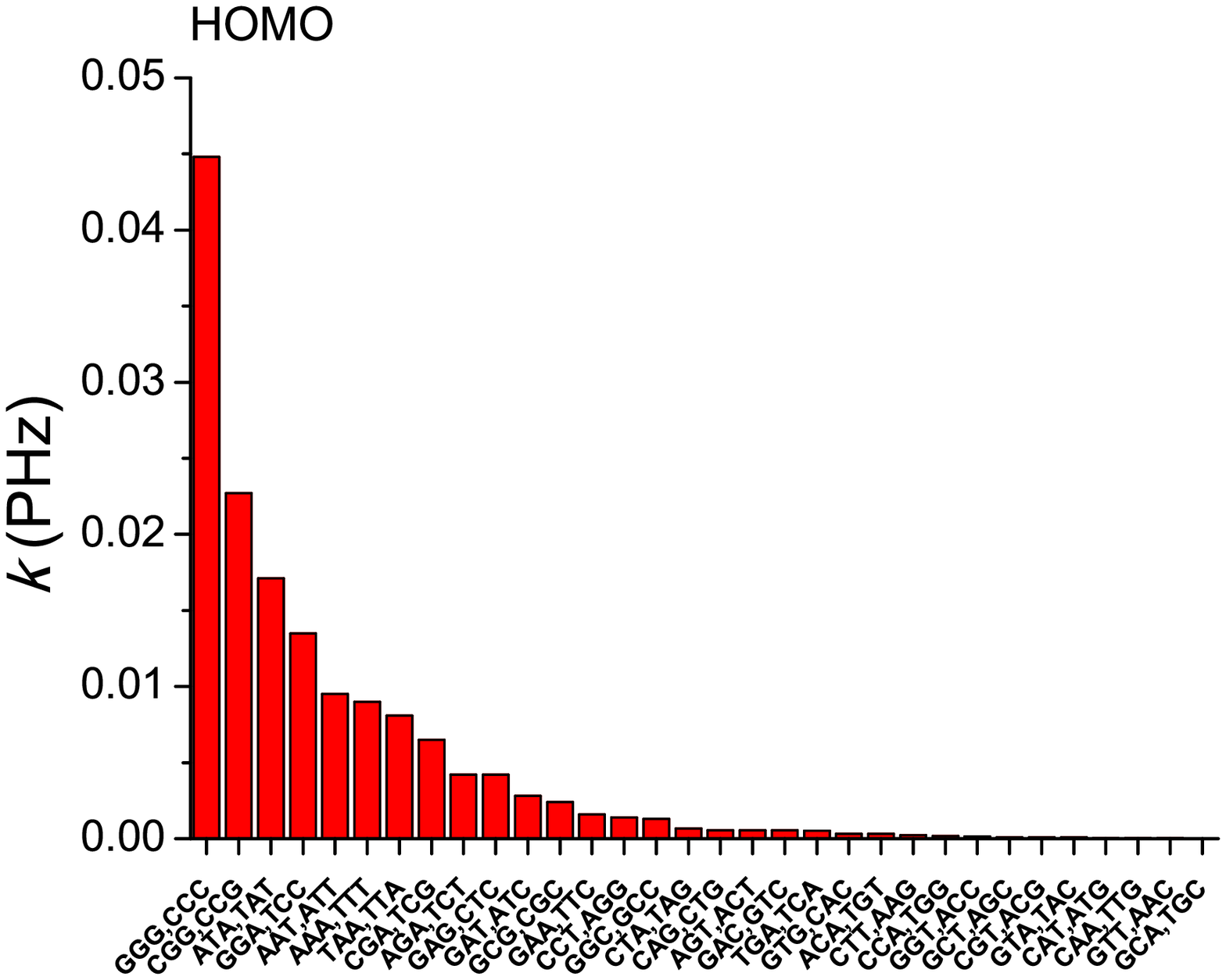}
\caption{HOMO \textit{pure} mean transfer rate $k$ for all trimers.}
\label{trimersholes}
\end{figure}
%For brevity, we will now focus on hole transfer.
%The calculations show that
Since for trimers consisting of different monomers $|A_{\mu}(t)|^2, \mu = 1, 2, 3$ may be non-periodic, from now on
we will only use the \textit{pure} mean transfer rate $k$, which if $|A_{3}(0)|^2 = 0$, can be defined as
%\begin{equation}\label{meantransferrate3}
$k = \frac{\langle |A_{3}(t)|^2 \rangle}{{t_{3}}_{mean}}$,
%\end{equation}
where ${t_{3}}_{mean}$ is the first time $|A_{3}(t)|^2$ becomes equal to $\langle |A_{3}(t)|^2 \rangle$
i.e. ``the mean transfer time''.
The HOMO \textit{pure} mean transfer rate $k$ for all possible trimers is shown in Fig.~\ref{trimersholes}.
For trimers consisting of identical monomers $k \approx 1.3109 \frac{p}{T}$.
As expected, $k$ is very small when trimers include dimers with very small $k$, primarily purines crossswise to pyrimidines
(GT \!\!\! $\equiv$ \!\!\! AC, CA \!\!\! $\equiv$ \!\!\! TG), secondarily AG \!\!\! $\equiv$ \!\!\! CT, thirdly GC.

For \emph{polymers}, supposing that $|A_{N}(0)|^2 = 0$,
for a polymer consisting of $N$ monomers, a \textit{pure} mean transfer rate can be defined as
%\begin{equation}\label{meantransferrateN}
$k = \frac{\langle |A_{N}(t)|^2 \rangle}{{t_{N}}_{mean}}$,
%\end{equation}
where ${t_{N}}_{mean}$ is the first time $|A_{N}(t)|^2$ becomes equal to $\langle |A_{N}(t)|^2 \rangle$ i.e.
``the mean transfer time''.
Increasing the number of base-pairs or monomers $N$, we study various characteristic polymers:
poly(dG)-poly(dC), poly(dA)-poly(dT), GCGCGC..., CGCGCG..., ATATAT..., TATATA...
as well as DNA segments that have been experimentally studied in the past.
If we fit $k(d)$ --i.e. the \textit{pure} mean transfer rate $k$ as a function of the charge transfer distance $d = N \times$ 3.4 {\AA}--
exponentially, as $k = k_0 \textrm{exp}(-\beta d)$, we obtain an estimation of $k_0$ and of the distance dependence parameter or inverse decay length $\beta$~\cite{Marcus}. These quantities are displayed in Table~\ref{table:ExponentialkofdFit}.
If, instead, we fit $k(N)$ --i.e. the \textit{pure} mean transfer rate $k$ as a function of the number of monomers $N$-- in a power law, as $k = k_0' N^{-\eta}$, we obtain an estimation of $k_0'$ and $\eta$.
These quantities are displayed in Table~\ref{table:PowerkofNFit}.
Values of $\beta$, in the range $\approx$ 0.3-1.5 {\AA}$^{-1}$, for various compounds, have been displayed in the literature at least 30 years now, see e.g Table IV of Ref.~\cite{Marcus}. In Table~\ref{table:ExponentialkofdFit} the values of $\beta$ are in the range $\approx$ 0.2-2 {\AA}$^{-1}$,
with smaller values for periodic polymers like ATATAT..., poly(dG)-poly(dC), poly(dA)-poly(dT). However, for efficient charge transfer, a small value of $\beta$ is not enough; one should also take into account the magnitude of $k_0$. The values of $k_0$ assumed in Ref.~\cite{Marcus} are 10$^{-2}$-10$^{-1}$ PHz which coincides with most of the $k_0$ values shown in Table~\ref{table:ExponentialkofdFit}, although generally, the values of $k_0$ fall in the wider range $\approx 10^{-4}$-10 PHz.
For the power law fit, $\eta \approx$ 1.7 - 17; most of the $k_0'$ values shown in Table~\ref{table:PowerkofNFit} are in the range $\approx 10^{-2}$-10$^{-1}$ PHz, although generally, the values of $k_0'$ fall in the wider range $\approx 10^{-4}$-10$^3$ PHz.
The $\beta$-value for charge transfer from an initial site (donor) to a final site (acceptor) depends on the mediating molecules, the so-called bridge. From Table~\ref{table:ExponentialkofdFit} we conclude that there are no universal values of $\beta$ and $k_0$ for DNA, instead, each specific DNA segment is unique and one should use an efficient and easy way to predict $\beta$ and $k_0$ of each DNA segment under investigation. It is hoped that the present  work will contribute in this direction.
$\beta$ values for different systems include
$\approx$ 1.0 - 1.4 {\AA}$^{-1}$ for protein-bridged systems %~\cite{WinklerGray:1992,Moser:1992,GrayWinkler:2005},
~\cite{Moser:1992,GrayWinkler:2005},
$\approx$ 1.55 - 1.65 {\AA}$^{-1}$ for aqueous glass bridges~\cite{GrayWinkler:2005},
$\approx$ 0.2 - 1.4 {\AA}$^{-1}$ for DNA segments~\cite{Lewis:1997,Holmlin:1998,Henderson:1999,Wan:2000,KawaiMajima:2013,KalosakasSpanou:2013},
$\approx$ 0.8 - 1.0 {\AA}$^{-1}$ for saturated hydrocarbon bridges~\cite{Johnson:1989,Oevering:1987},
$\approx$ 0.2 - 0.6 {\AA}$^{-1}$ for unsaturated phenylene~\cite{Helms:1992,Ribou:1994},
%polyene~\cite{Arrhenius:1986,Effenberger:1991,Tolbert:1992}
polyene~\cite{Effenberger:1991,Tolbert:1992}
and %polyyne~\cite{Tour:1996,Grosshenny:1996,Sachs:1997}
polyyne~\cite{Grosshenny:1996,Sachs:1997}
bridges, and much smaller values ($<$ 0.05 {\AA}$^{-1}$), suggesting a molecular-wire-like behavior, for a p-phenylenevinylene bridge~\cite{Davis:1998}.
Hence, it seems that charge transfer in ATATAT..., poly(dG)-poly(dC) and poly(dA)-poly(dT) is almost molecular-wire-like.
Since a carrier can migrate along DNA over 200 {\AA}~\cite{Meggers:1998,Henderson:1999,KawaiMajima:2013},
in the present calculations for polymers $d$ is extending up to 204 {\AA} ($N$ up to 60 base-pairs).

%\begin{widetext}

\begin{table}[h!]
\caption{$k_0$ and $\beta$ of the exponential fit $k = k_0 \textrm{exp}(-\beta d)$ for various DNA polymers.
C.C. is the correlation coefficient.}
\scriptsize{
\begin{tabular}{|l|c|c|c|c|} \hline
DNA segment       & $k_0$ (PHz)       & $\beta$ ({\AA}$^{-1}$) & C.C.        &H/L\\
                  &                   &                        &             &   \\ \hline
poly(dG)-poly(dC) & 0.176 $\pm$ 0.007 & 0.189 $\pm$ 0.008      & 0.988       & H \\ \hline
poly(dG)-poly(dC) & 0.035 $\pm$ 0.001 & 0.189 $\pm$ 0.007      & 0.989       & L \\ \hline
poly(dA)-poly(dT) & 0.035 $\pm$ 0.001 & 0.189 $\pm$ 0.008      & 0.988       & H \\ \hline
poly(dA)-poly(dT) & 0.051 $\pm$ 0.002 & 0.189 $\pm$ 0.008      & 0.989       & L \\ \hline
GCGCGC...         & 0.032 $\pm$ 0.003 & 0.358 $\pm$ 0.023      & 0.988       & H \\ \hline
ATATAT...         & 0.057 $\pm$ 0.002 & 0.168 $\pm$ 0.008      & 0.985       & H \\ \hline
CGCGCG...         & 0.932 $\pm$ 0.233 & 0.871 $\pm$ 0.074      & 0.994       & H \\ \hline
TATATA...         & 0.110 $\pm$ 0.005 & 0.251 $\pm$ 0.012      & 0.985       & H \\ \hline
AGTGCCAAGCTTGCA   & 0.059 $\pm$ 0.002 & 0.685 $\pm$ 0.008      & 1.000       & H \\ \hline
AGTGCCAAGCTTGCA   & $(9.8 \! \pm \! 2.6) \!\! \times \!\! 10^{-5}$& 0.197 $\pm$ 0.059      & 0.808       & L \\ \hline
TAGAGGTGTTATGA    & 4.306 $\pm$ 5.001 & 1.321 $\pm$ 0.342      & 0.998       & H \\ \hline
TAGAGGTGTTATGA    & 2.877 $\pm$ 0.833 & 2.154 $\pm$ 0.085      & 1.000       & L \\ \hline
\end{tabular} }   \label{table:ExponentialkofdFit}
\end{table}

\begin{table}[h!]
\caption{$k_0'$ and $\eta$ of the power fit $k = k_0' N^{-\eta}$ for various DNA polymers.
C.C. is the correlation coefficient.}
\scriptsize{
\begin{tabular}{|l|c|c|c|c|} \hline			
DNA segment       & $k_0'$  (PHz)     & $\eta$            & C.C.        &H/L\\
                  &                   &                   &             &   \\ \hline
poly(dG)-poly(dC) & 0.359 $\pm$ 0.001 & 1.893 $\pm$ 0.002 & 1.000       & H \\ \hline
poly(dG)-poly(dC) & 0.072 $\pm$ 0.000 & 1.895 $\pm$ 0.002 & 1.000       & L \\ \hline
poly(dA)-poly(dT) & 0.072 $\pm$ 0.000 & 1.892 $\pm$ 0.002 & 1.000       & H \\ \hline
poly(dA)-poly(dT) & 0.105 $\pm$ 0.000 & 1.893 $\pm$ 0.002 & 1.000       & L \\ \hline
GCGCGC...         & 0.087 $\pm$ 0.008 & 3.176 $\pm$ 0.127 & 0.993       & H \\ \hline
ATATAT...         & 0.117 $\pm$ 0.004 & 1.776 $\pm$ 0.035 & 0.994       & H \\ \hline
CGCGCG...         & 5.082 $\pm$ 1.619 & 6.715 $\pm$ 0.458 & 0.994       & H \\ \hline
TATATA...         & 0.236 $\pm$ 0.007 & 2.295 $\pm$ 0.035 & 0.997       & H \\ \hline
AGTGCCAAGCTTGCA   & 1.383 $\pm$ 0.826 & 4.487 $\pm$ 0.487 & 0.997       & H \\ \hline
AGTGCCAAGCTTGCA   & $(2.2 \! \pm \! 1.0)\!\! \times \!\! 10^{-4}$ & 2.176 $\pm$ 0.543 & 0.761       & L \\ \hline
TAGAGGTGTTATGA    &46.300 $\pm$53.288 & 9.902 $\pm$ 1.660 & 0.998       & H \\ \hline
TAGAGGTGTTATGA    &203.457$\pm$99.552& 16.708 $\pm$ 0.706& 1.000       & L \\ \hline
\end{tabular} }   \label{table:PowerkofNFit}
\end{table}

%\end{widetext}

In Ref.~\cite{WangLewisSankey:2004} the authors calculated the complex band structure of poly(dA)-poly(dT) and poly(dG)-poly(dC) using an \textit{ab initio} tight-binding method based on density-functional theory and obtained the energy dependence $\beta(E)$.
%They also found the bang gap of poly(dA)-poly(dT) (2.7 eV) and of poly(dG)-poly(dC) (2.0 eV) in accordance with previous works.
%Within the fundamental band gap between valence and conduction band i.e. HOMO and LUMO in our approach, they found several $\beta(E)$ curves.
Since the states with large $\beta$ values don't play a significant role in conduction they noticed that only the smallest $\beta(E)$ states, described by a semielliptical-like curve in the band-gap region are important. This branch reaches a maximum $\beta$ value near midgap, called the branch point, $\beta_{bp}$, $\approx$ 1.5 {\AA}$^{-1}$ both for poly(dA)-poly(dT) and poly(dG)-poly(dC). Since in molecular electronics metallic contacts are made at the two ends of the molecule and electronic current is carried by electrons tunneling from the metal with energies in the band-gap region, the branch point plays an important role in the conductance. Although the above hold when metal conducts are attached to the molecule, in photoinduced charge transfer experiments, we are interested in states close to the top of the valence band i.e. the HOMO or close to the bottom of the conduction band i.e. the LUMO.
%As far as one can judge from the figures,
For the top of the valence band of poly(dA)-poly(dT) [Fig.1a of Ref.~\cite{WangLewisSankey:2004}] $\beta \approx $ 0.4 {\AA}$^{-1}$ and for poly(dG)-poly(dC) [Fig.1b of Ref.~\cite{WangLewisSankey:2004}] $\beta \approx $ 0.2 {\AA}$^{-1}$,
%These values are
close to the values predicted in the present work
($\approx$ 0.2 {\AA}$^{-1}$ both for poly(dA)-poly(dT) and poly(dG)-poly(dC) cf. Table~\ref{table:ExponentialkofdFit}).
%For the bottom of the conduction band of poly(dA)-poly(dT) [Fig.1a of Ref.~\cite{WangLewisSankey:2004}] $\beta \approx $ 0.5 {\AA}$^{-1}$ and for poly(dG)-poly(dC) [fig.1b of Ref.~\cite{WangLewisSankey:2004}] $\beta \approx $ 0.7 {\AA}$^{-1}$.
%On the contrary, the values predicted in the present work are $\approx$ 0.2 {\AA}$^{-1}$ both for poly(dA)-poly(dT) and poly(dG)-poly(dC) cf. Table~\ref{table:ExponentialkofdFit}.
%Maybe this difference is related to the fact that in Ref.~\cite{WangLewisSankey:2004}, the complex band structure is calculated using an \textit{ab initio} tight-binding method based on density-functional theory, while, from our point of view, we expect no difference in the $\beta$ both for poly(dA)-poly(dT) and poly(dG)-poly(dC) and both for the HOMO or the LUMO since all these are the same type of polymers with just a different strength of interaction between the constituting monomers (hence the $k_0$ values are different and proportional to the corresponding hopping parameters).

\begin{figure}[h!]
\hspace{0cm}
\centering
\includegraphics[width=7cm]{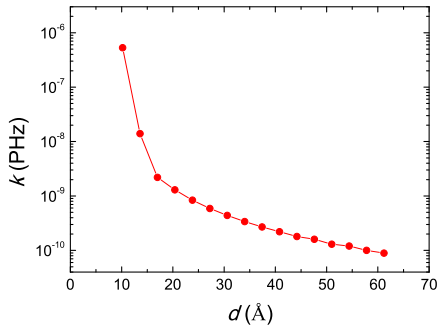}
\caption{Experiment of Giese \textit{et al.}~\cite{Giese:2001}, i.e. [G] (T)n [GGG] TATTATATTACGC.
(T)n denotes the bridge made up from $n$ T-A monomers between
 the hole donor G (the first G-C monomer) and the hole acceptor GGG (the trimer made by three G-C monomers) before the TATTATATTACGC tail.
the hole donor and acceptor denoted by square brackets.
For the strong distance dependence $k(d)$ range (for $n = $ 1,2,3), $\beta \approx$ 0.8 {\AA}$^{-1}$.
In the experiment [Figure 3 of Ref.~\cite{Giese:2001}] the authors find qualitatively the same behavior,
while they estimate $\beta \approx$ 0.6 {\AA}$^{-1}$ (for $n = 1, 2, 3$) i.e. for the strong distance dependence range.
For the weak distance dependence region, again in agreement with the experiment,
a much weaker distance dependence with $\beta \approx$ 0.07 {\AA}$^{-1}$ is obtained.
}
\label{fig:Giese2001kofd}
\end{figure}

In Ref.~\cite{Giese:2001} Giese \textit{et al.} studied experimentally the hole transfer in the DNA segment
[G] (T)n [GGG] TATTATATTACGC. (T)n denotes the bridge made up from $n$ T-A monomers between
the hole donor [G] %(the first G-C monomer)
and the hole acceptor [GGG]
%(the trimer made by G-C monomers)
%The hole donor and acceptor are
denoted by square brackets,
before the TATTATATTACGC tail.
In Fig.~\ref{fig:Giese2001kofd} the computed $k(d)$ i.e. the \textit{pure} mean transfer rate as a function of the distance from the hole donor to the middle of the hole acceptor is shown.
In accordance with the experiment~\cite{Giese:2001} we find two regions with different distance dependence.
For $n = 1, 2, 3$ the distance dependence is strong becoming much weaker for $n \ge 4$.
For the strong distance dependence range, we find $\beta \approx$ 0.8 {\AA}$^{-1}$.
In the experiment [Fig. 3 of Ref.~\cite{Giese:2001}] the authors find qualitatively the same behavior, estimating $\beta \approx$ 0.6 {\AA}$^{-1}$ for $n = 1, 2, 3$.
For $n =4, \dots , 16$ we compute a much weaker distance dependence with $\beta \approx$ 0.07 {\AA}$^{-1}$.
%Hence, the approach presented here explains the experimental results of Ref.~\cite{Giese:2001}.
%both qualitatively and quantitatively
%Moreover, we obtain almost identical results if we exclude the TATTATATTACGC tail (not shown here).
%While this manuscript was under construction Ref.~\cite{KalosakasSpanou:2013} was published where the authors, using a description of charge transfer at the single base level~\cite{HKS:2010-2011}, a different estimation of the mean transfer time involving Monte Carlo simulations [in contrast, in the present article the mean transfer time to the $j$ monomer ${t_{j}}_{mean}$ is the first time $|A_{j}(t)|^2$ becomes equal to $\langle |A_{j}(t)|^2 \rangle$], a different definition of the transfer rates being in Ref.~\cite{KalosakasSpanou:2013} just the inverse of the mean transfer time [in contrast, in the present article it is given by e.g. Eq.~\ref{meantransferrateN}, including $\langle |A_{j}(t)|^2 \rangle$], and somehow different tight-binding parameters, manage to explain qualitatively the experimental results of Ref.~\cite{Giese:2001}. However the authors notice ``a numerical discrepancy between the experimental and the calculated results is that the switching of the behavior that is obtained for $n > $ 4 in Fig. 1 is observed for $n > $ 3 in ref. 11'' and ``we see that the numerical results within our model overestimate the experimental value of $\beta$ by a factor of 2''.

In Ref.~\cite{Murphy:1993} the authors demonstrated rapid photoinduced electron transfer over a distance of greater than
40 {\AA} between metallointercalators tethered to the 5$'$ termini of %a 15-base-pair DNA duplex (AGTGCCAAGCTTGCA).
AGTGCCAAGCTTGCA.
The authors~\cite{Murphy:1993} mentioned that ``the photoinduced electron transfer between intercalators occurs very rapidly over $>$ 40 {\AA} through the DNA helix over a pathway consisting of $\pi$-stacked base-pairs.'' Then, from Marcus theory \cite{Marcus}
they estimated $\beta$ to be $\leq$ 0.2 {\AA}$^{-1}$. We observe (Table~\ref{table:ExponentialkofdFit}) that for electron transfer (through LUMOs) we also find $\beta \leq$ 0.2 {\AA}$^{-1}$, while for hole transfer (through HOMOs) we find $\beta \approx$ 0.7 {\AA}$^{-1}$.
Similar weak distance dependence with $\beta \leq$ 0.2 {\AA}$^{-1}$ was found in Ref.~\cite{Arkin:1996}.

\begin{figure}[]
\hspace{0cm}
\centering
\includegraphics[width=7cm]{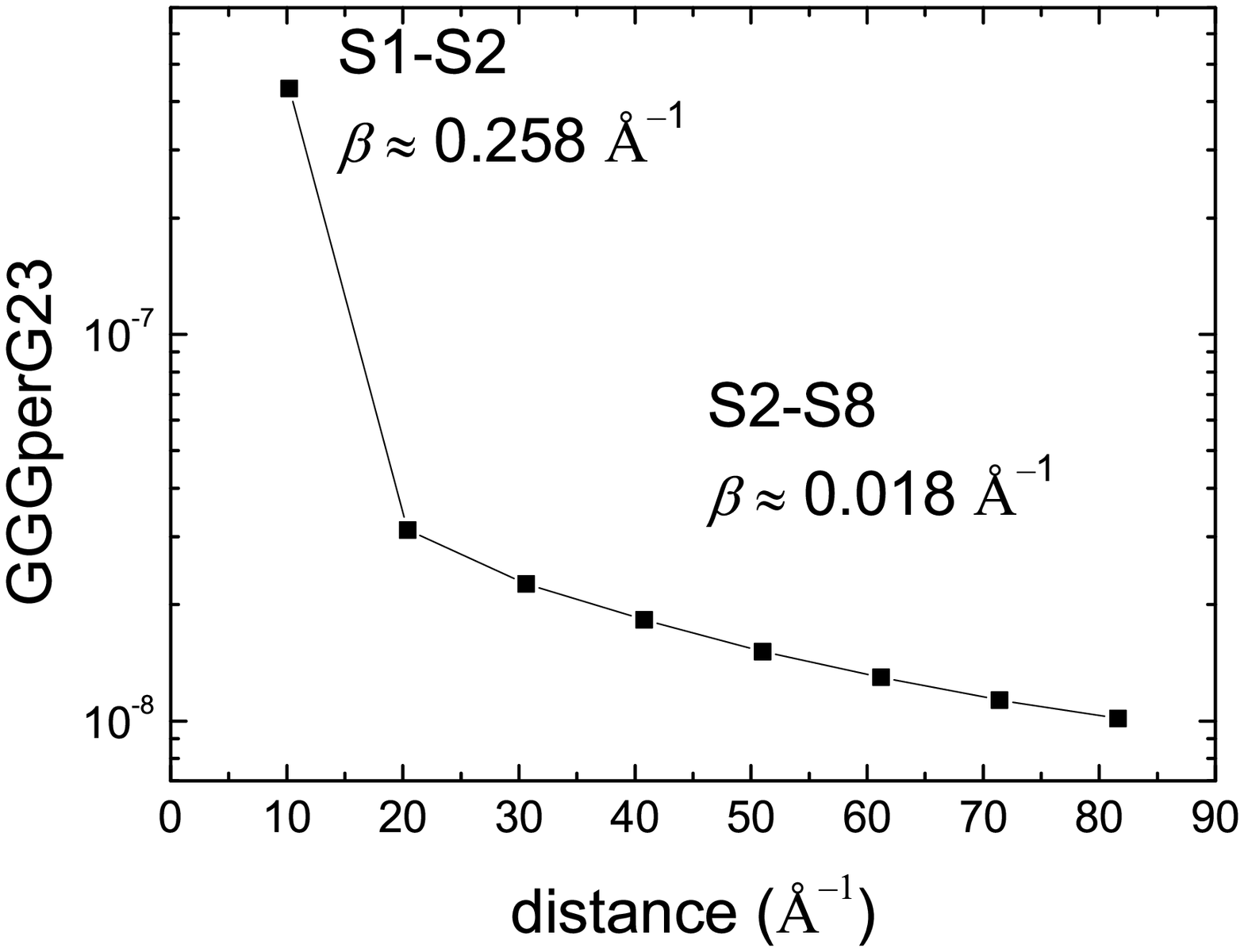}
\includegraphics[width=7cm]{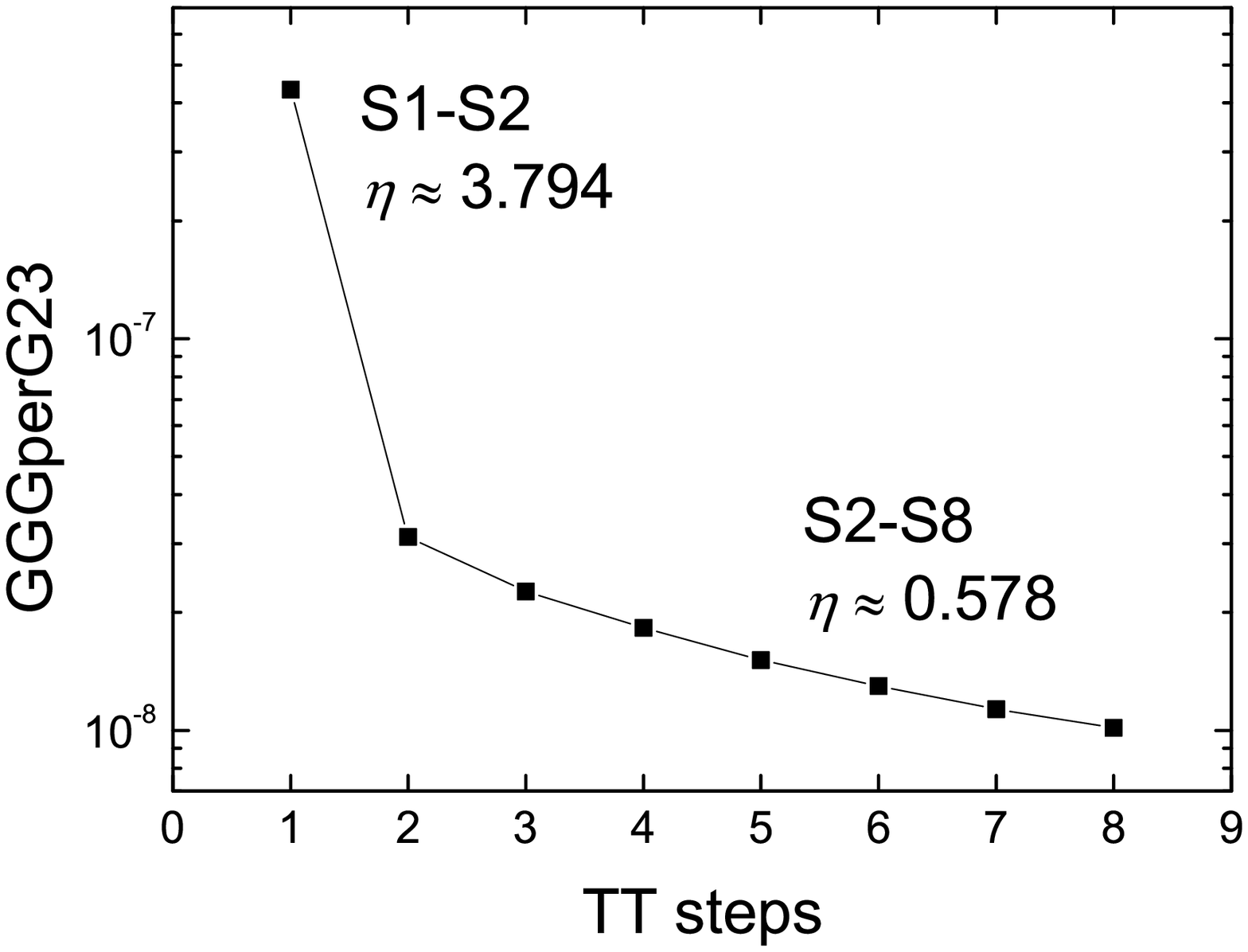}
\caption{Hole transfer in ACGCACGTCGCATAATATTACG [bridge] GGGTATTATATTACGC.
The [bridge] is made up of TT dimers separated by G monomers. In the experiment~\cite{Giese:1999}, [bridge] is either TT (one TT step) either TTGTT (two TT steps) or TTGTTGTTGTT (four TT steps).
%The hole is created in the C-G monomer before the G-C monomer before the [bridge], and transferred to the GGG trimer. Following Ref.~\cite{Giese:1999}, the distance is measured from G23 to the ``left'' G of GGG.
%$k_{\textrm{leftG}}$, $k_{\textrm{middelG}}$, and $k_{\textrm{rightG}}$ refer to the \textit{pure} mean transfer rate of the first, second, and third G of GGG.
}
\label{fig:Giese1999}
\end{figure}

In Ref.~\cite{Giese:1999} the authors study hole transfer in the DNA sequence
ACGCACGTCGCATAATATTACG [bridge] GGGTATTATATTACGC,
where the [bridge] is either TT (sample 1a, one TT step) either TTGTT (sample 2a, two TT steps) or TTGTTGTTGTT (sample 3a, four TT steps).
The hole is created in the C-G monomer before the G-C monomer before the [bridge] and transferred to the GGG trimer. The charge transfer is measured by ``the oxidative damage at the G and GGG units'', ``quantified after piperidine
treatment and polyacrylamide gel electrophoresis with a phospho-imager''. To compare our results with the experiment we need the ratio
of $\sum_j \langle |A_{j}(t)|^2 \rangle$ where $j$ represents the three monomers of the GGG trimer
to $\langle |A_{i}(t)|^2 \rangle$ where $i$ represents the initial G-C monomer (called also G$_{23}$).
This ratio is called GGGperG23 in Fig.~\ref{fig:Giese1999}.
%A similar comparison has been published as this manuscript was still under construction in Ref.~\cite{KalosakasSpanou:2013}, where the authors, using a description of charge transfer at the single base level~\cite{HKS:2010-2011}, a different estimation of the mean transfer time involving Monte Carlo simulations [in contrast, in the present article the mean transfer time to the $j$ monomer ${t_{j}}_{mean}$ is the first time $|A_{j}(t)|^2$ becomes equal to $\langle |A_{j}(t)|^2 \rangle$],a different definition of the transfer rates being in Ref.~\cite{KalosakasSpanou:2013} just the inverse of the mean transfer time [in contrast, in the present article it is given by e.g. Eq.~\ref{meantransferrateN}, including $\langle |A_{j}(t)|^2 \rangle$], and somehow different tight-binding parameters, manage to explain qualitatively the experimental results of Ref.~\cite{Giese:1999}.
Our calculations with three or four TT steps confirm the experiment either using an exponential fit with the $\beta$ parameter or a power law fit with the $\eta$ parameter.
Extending the present approach up to eight TT steps reveals (Fig.~\ref{fig:Giese1999}) that
there are two distinct regions (i) one step (S1) to two steps (S2), and (ii) more than two steps (up to eight steps are included in the graphs).
%Following Ref.~\cite{Giese:1999}, in Fig.~\ref{fig:Giese1999} the distance is measured from G$_{23}$ to the ``left'' G of GGG. In Fig.~\ref{fig:Giese1999}, $k_{\textrm{leftG}}$, $k_{\textrm{middelG}}$, and $k_{\textrm{rightG}}$ refer to the \textit{pure} mean transfer rate of the first, second, and third G of GGG.

A handy method to examine the charge transfer properties of DNA segments was displayed.
%It allows to illustrate to which extent a specific DNA segment can serve as an efficient medium for charge transfer. The temporal and spatial evolution of electrons or holes along a $N$ base-pair DNA segment can be determined, solving a system of $N$ coupled differential equations. As input one needs the relevant on-site energies of the base-pairs and the hopping parameters between successive base-pairs.
%The method can be applied to any DNA segment. In the present manuscript, it has been applied to all possible dimers either for holes or for electrons, to all possible trimers for holes, and to various polymers (e.g.  poly(dG)-poly(dC), poly(dA)-poly(dT), GCGCGC..., ATATAT...) either for electrons or holes. The results have been succesfully compared with results obtained by other workers, especially experimental ones.
Useful physical quantities were obtained including the \textit{pure} mean carrier transfer rate $k$,
the inverse decay length $\beta$ used for an exponential fit ($k = k_0 \textrm{exp}(-\beta d)$) of the transfer rate as a function of the charge transfer distance $d = N \times$ 3.4 {\AA} and the exponent $\eta$ used for a power law fit ($k = k_0' N^{-\eta}$) of the transfer rate as function of the number of monomers $N$.
The values of these parameters are not universal, depend on the specific DNA segment and are different for electrons and holes.
%Approximately, $\beta$ falls in the range $\approx$ 0.2 - 2 {\AA}$^{-1}$, $k_0$ is usually 10$^{-2}$-10$^{-1}$ PHz although, generally, it falls in the wider range 10$^{-4}$-10 PHz, while $\eta$ falls in the range $\approx$ 1.7 - 17, $k_0'$ is usually $\approx 10^{-2}$-10$^{-1}$ PHz, although generally, it falls in the wider range $\approx 10^{-4}$-10$^3$ PHz.

%\begin{acknowledgments}
%The author wishes to thank ...
%\end{acknowledgments}

\end{document}